# A Phoneme-Scale Assessment of Multichannel Speech Enhancement Algorithms

Nasser-Eddine Monir[1], Paul Magron[1], Romain Serizel[2]

Université de Lorraine, CNRS, Inria, Loria

[1]{nasser-eddine.monir, paul.magron}@inria.fr

[2]romain.serizel@loria.fr

# Abstract

In the intricate acoustic landscapes where speech intelligibility is challenged by noise and reverberation, multichannel speech enhancement emerges as a promising solution for individuals with hearing loss. Such algorithms are commonly evaluated at the utterance level. However, this approach overlooks the granular acoustic nuances revealed by phoneme-specific analysis, potentially obscuring key insights into their performance. This paper presents an in-depth phoneme-scale evaluation of 3 state-of-the-art multichannel speech enhancement algorithms. These algorithms -FasNet, MVDR, and Tango- are extensively evaluated across different noise conditions and spatial setups, employing realistic acoustic simulations with measured room impulse responses, and leveraging diversity offered by multiple microphones in a binaural hearing setup. The study emphasizes the fine-grained phoneme-level analysis, revealing that while some phonemes like plosives are heavily impacted by environmental acoustics and challenging to deal with by the algorithms, others like nasals and sibilants see substantial improvements after enhancement. These investigations demonstrate important improvements in phoneme clarity in noisy conditions, with insights that could drive the development of more personalized and phoneme-aware hearing aid technologies.

# Introduction

Speech within a noisy environment is a complicated scenario that can substantially diminish the clarity of spoken words. Background noise can obscure important acoustic cues, challenging listeners in differentiating individual speech sounds and words. Speech enhancement is a solution to enhance speech intelligibility in noisy environments (Loizou, 2007). This technique estimates the speech signal from the noisy mixture, by relying on acoustic cues and temporal patterns inherent to the speech. Speech enhancement algorithms are broadly categorized into two types: single-channel and multichannel, depending on the number of available microphones to record the sound.

In scenarios where audio is captured through a single microphone, speech enhancement algorithms concentrate on temporal, frequency, and spectro-temporal characteristics to filter out the noise (Loizou, 2007). Such single-channel speech enhancement is limited to the information captured by one reference point, and often focuses on aspects like noise variance over time or spectral consistency.

Conversely, multichannel speech enhancement algorithms harness the power of spatial diversity by exploiting the various captures of speech across microphones (Benesty, Chen, & Huang, 2008). This multifold capture allows for considering the spatial characteristics and the directionality of sound. By comparing the different signal channels obtained at the microphones, these algorithms offer a more robust reconstruction of the original speech, effectively mitigating the masking effects of background noise.

As a subset of multichannel speech enhancement algorithms, beamformers manipulate spatial sound attributes using microphone arrays (Benesty, Chen, & Huang, 2008). Unlike broader multichannel methods that filter or cancel noise, beamformers enhance speech intelligibility by precisely manipulating the spatial attributes of the acoustic signal (sound's directionality). They amplify speech from a specific direction while reducing noise and reverberation from others. This targeted approach is particularly effective in noisy environments, where it isolates the speaker's voice from disruptive background sounds, enhancing speech clarity and intelligibility.

Current state-of-the-art multichannel speech enhancement systems are characterized by advanced beamforming techniques and the integration of neural networks to improve the intelligibility of speech in noise. The minimum variance distortionless response (MVDR) beamformer optimizes noise reduction while preserving the desired speech directionality (Capon, 1969) (Heymann, Drude, & Haeb-Umbach, 2016). In hybrid methods, neural networks provide parameters for signal processing filters. These include the distributed multichannel Wiener filter (MWF) (Bertrand & Moonen, 2010) (Furnon, Serizel, Essid, & Illina, 2021) and its adaptations like the generalized eigenvalue decomposition MWF (GEVD-MWF) (Serizel, Moonen, Van Dijk, & Wouters, 2014). Alternatively, some approaches relying entirely on neural networks have been proposed, such as the filter-and-sum network (FaSNet) beamformer (Luo, Han, Mesgarani, Ceolini, & Liu, 2019). This model uses neural networks to directly predict signals rather than the parameters of a spatial filter, which allows for enhanced flexibility in optimization. These developments reflect a shift towards sophisticated, binaural processing setups where hearing aids on both sides collaborate, leveraging spatial information to differentiate speech from noise effectively (Kollmeier & Koch, 1994) (Van den Bogaert, Doclo, Wouters, & Moonen, 2009).

Usually, speech enhancement algorithms are evaluated at the *utterance* level using objective signal-to-noise ratio-like metrics, which offers a convenient way to quantify their performance at a coarse level and compare algorithms. However, this method does not capture the nuanced ways different phonemes react to noise, nor the way algorithms process these phonemes, which potentially simplify their true effectiveness. Studies contrasting English phoneme recognition in noise for native and non-native speakers reveal this complexity (Adachi, Akahane-Yamada, & Ueda, 2006). For instance, Miller and Nicely. indicates that consonants vary in noise tolerance, suggesting that some phonemes are more susceptible to noise masking than others (Miller & Nicely, 1955). This variance may significantly affect the perceived effectiveness of speech enhancement models. Furthermore, phoneme confusion observed in both human and automatic speech recognition systems suggests that consonants and vowels experience different levels of impact from information loss due to noise (Meyer, Jürgens, Wesker, Brand, & Kollmeier, 2010) (Zaar & Dau, 2017). Studies have shown that a degraded classification of voicing can lead to more confusion between voiced and unvoiced phonemes, such as /p/ and /b/, whereas phonemes differing in the place of articulation, like /p/ and /d/, remain distinguishable (Dubno & Levitt, 1981) (Gelfand, Piper, & Silman, 1985). This distinct impact of noise on phoneme representation in auditory and speech motor systems suggests that evaluating speech enhancement models should consider these phonetic and acoustic variations. Furthermore, the age and cognitive factors in speech perception, as well as the speech-intrinsic variability such as speaking style, can influence phoneme recognition in noise. These factors underline the complexity of real-world hearing scenarios and the necessity of a more detailed, phoneme-level analysis in assessing speech enhancement models. Without accounting for these diverse acoustic responses of phonemes to noise, there is a risk of underestimating a model's practical performance in real-life scenarios.

In this paper, we propose to evaluate three state-of-the-art speech enhancement algorithms at a phoneme-level for a nuanced analysis that aligns with the distinct acoustic properties of phonetic elements. Such detailed scrutiny can reveal the specific strengths and weaknesses of models in preserving the fidelity of speech sounds, offering valuable insights for tailoring speech enhancement techniques to the challenges posed by different types of noise.

By analyzing the performance of FasNet (Luo, Han, Mesgarani, Ceolini, & Liu, 2019), a neural network based MVDR (Heymann, Drude, & Haeb-Umbach, 2016), and Tango (a distributed MWF) (Furnon, Serizel, Essid, & Illina, 2021), this study goes beyond identifying the best model. Instead, it focuses on refining evaluation methodologies at the phoneme level. By closely examining how each model handles the clarity of individual phonemes amidst different noise types, we uncover the nuanced strengths and weaknesses of each approach. This detailed phoneme-level evaluation reveals how these models interact with the unique acoustic characteristics of specific phonemes. The goal of this comparative analysis is to guide the development of speech enhancement algorithms, ensuring they are tuned to enhance phonemic clarity across the varied acoustic environments encountered in everyday listening scenarios.

The rest of this paper is structured as follows. First, we provide an overview of multichannel speech enhancement by setting the problem and detailing the algorithms we use in our study. Then, the methodology section delves into the process of data collection and generation, and notably highlights the phoneme classification. The next section describes our extensive experiments and discusses its results, with a particular emphasis on the phoneme-scale evaluation. Finally, the last section draws some concluding remarks.

## Overview of Multichannel Speech Enhancement

### Problem Statement and Notations

Let us consider an acoustic scenario with two punctual sources and several microphones distant from the sources. One source is considered as the target speech source, while the other is an interfering source. In the case of hearing aids, which is of interest here, we have $M$ microphones on each hearing aid and two hearing aids: one on the left side, and one on the right side, respectively denoted $L$ and $R$ in the remainder of the paper. This scenario is illustrated in Figure 1.

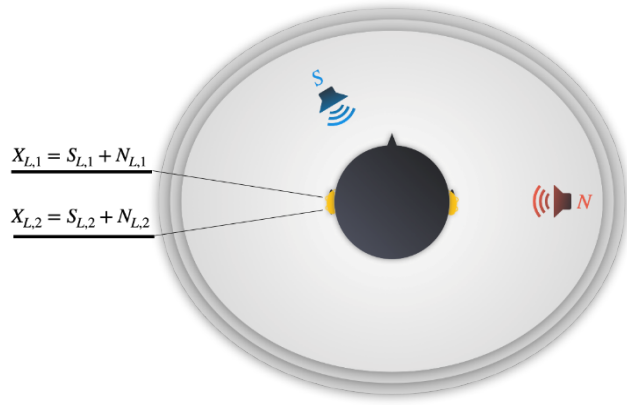

*Figure 1 - Spatialized acoustic scenarios with two sources: a speech source and a noise source. The acoustic sources are point sources. The signal $X_{i,m}$ recorded by each of the microphone is the sum of the reverberated images of these sources $S_{i,m}$ and $N_{i,m}$ at the hearing aids microphones.*

We note $s$ and $n \in \mathbb{R}^{\tilde{T}}$ the time-domain speech and noise signals, respectively, where $\tilde{T}$ denotes the length (in samples) of these signals. We assume that the signals emitted by the source are band-limited and we express these signals in the time-frequency domain by means of the short-time Fourier transform. The signal emitted by the target speech source is denoted $S \in \mathbb{C}^{T \times F}$, where $T$ and $F$ are the number of time frames and frequency channels, respectively. The $(t, f) -$ th entry of this signal is denoted $S(t, f)$. Similarly, the signal emitted by the interfering source is denoted $N$. The contribution of the speech signal recorded at the $m^{\text{th}}$ microphone of the right (resp. left) hearing aids is denoted $S_{R,m}$ (resp. $S_{L,m}$). Similarly, $N_{R,m}$ and $N_{L,m}$ denote the contribution of the interfering noise source at the $m^{\text{th}}$ microphone of the right and left hearing aid, respectively. These signals are commonly referred to as *images* of the speech (resp. the noise) at the microphones.

The noisy mixture signal observed at the hearing aids is the combination of the speech and source images:

$$X_{i,m} = S_{i,m} + N_{i,m}, \ i \in \{L, R\}.$$

Finally, we note $\mathbf{X}_R = [X_{R,1}, \dots, X_{R,M}]$ the set of noisy signals recorded at the right hearing aid, and similarly for $\mathbf{X}_L$, and we note $\mathbf{X}_{\text{Bin}}$ the set of all mixture signals (the same notation can be applied to the target speech and noise signals as well).

Speech enhancement is a general term that encompasses noise reduction (Loizou, 2007), dereverberation (Naylor & Gaubitch, 2010) or a combination of both. The goal of noise reduction is to estimate the speech signal at an arbitrary microphone $S_{i,m}$ given the recorded mixture $\mathbf{X}_i \ i \in \{L, R, \text{Bin}\}$. The goal of dereverberation is to estimate the speech source $S$ from the speech recorded by one or several microphone $S_{i,m}$. Estimating the speech source $S$ from the recorded mixture $\mathbf{X}_i$ is also considered as speech enhancement and could be seen as a combination of noise reduction and dereverberation. In this paper we will focus on noise reduction.

*In hearing aid scenarios that we consider here, there are two possible speech enhancement setups, illustrated in Figure 1*

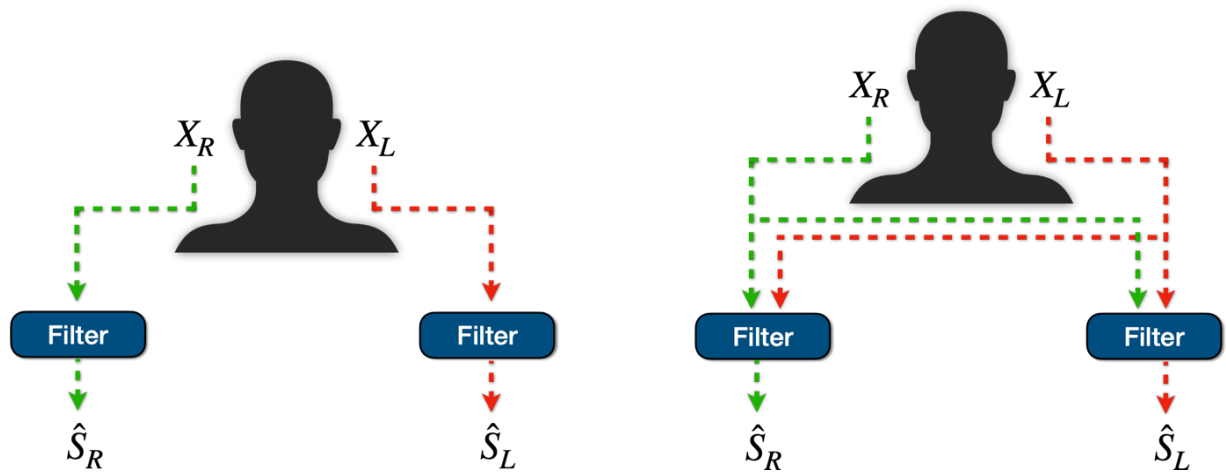

Figure 2. On the one hand, in the *bilateral* setup there is no communication between the hearing aids and each side is processed independently. Therefore, the input of the left hearing aid filter is $\mathbf{X}_L$, and the input of the right hearing aid filter is $\mathbf{X}_R$. Each hearing aid can then be seen as a compact microphone antenna (Benesty, Chen, & Huang, 2008) where each microphone relative position is fixed.

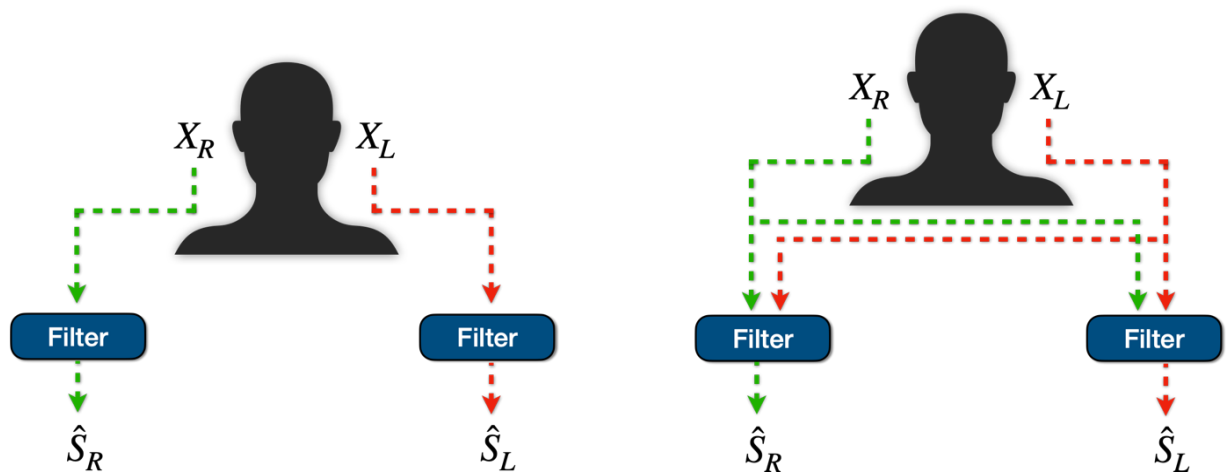

*Figure 2 – Speech enhancement setups: bilateral (left) and binaural (right).*

On the other hand, in the *binaural* speech enhancement setup (Kollmeier & Koch, 1994), the hearing aids can communicate with each other, thus each filter has access to more diverse information about the acoustic scene. This can improve the filter's effectiveness, notably in asymmetric acoustic scenarios (Van den Bogaert, Doclo, Wouters, & Moonen, 2009). This scenario is also more generally referred to as distributed microphone arrays (Bertrand & Moonen, 2010). In this setup, each hearing aid filter processes the whole set of noisy signals $\mathbf{X}_{\text{Bin}}$ as input, but the input signals at the left and the right filters are ordered differently

($\mathbf{X}_{\text{Bin},R} = [\mathbf{X}_R, \mathbf{X}_L]$ for the right filter vs. $\mathbf{X}_{\text{Bin},L} = [\mathbf{X}_L, \mathbf{X}_R]$ for the left one). This ordering implies that each filter uses a different reference signal. The reference signal is usually selected as one of the input signals from the ear of interest (i.e., not a channel from the contralateral ear). In the binaural setup, the filters have access to the signal collected from a combination of compact microphones array. In the remainder of the paper, we will focus on the binaural setup. We will also assume that the signal transmission between the hearing aids is *perfect*, i.e., it does not introduce any latency of packet loss. The impact of the communication channel between hearing aids as encountered in a real device is beyond the scope of the current paper.

## Methods

As opposed to single-channel filtering that relies on a single microphone signal (Loizou, 2007), multichannel filters can exploit spatial information about the acoustic scene that is obtained through the signals recorded by the multiple microphones, as detailed above (Benesty, Chen, & Huang, 2008). Because of their ability to focus on one direction in space, these approaches are commonly referred to as *beamformers*. For the past years, multichannel speech enhancement relying partly or totally on neural networks has experienced significant performance improvement and better applicability of the filters in realistic scenarios. Two main strategies are used to exploit neural networks for multichannel speech enhancement.

The first set of approaches leverages traditional signal processing techniques to design spatial filters. Considering a given cost function to minimize, a filter $\mathbf{W} \in \mathbb{C}^M$ is obtained as the closed form solution to this optimization problem for each time-frequency point. This filter can therefore be applied to the noisy mixture to yield the speech estimate:

$$\hat{S}_{i,m}(f,t) = W^H(f,t) X_i(f,t), \; i \in \{L, R, Bin\},$$

where $\cdot^H$ denotes the Hermitian transpose.

Neural networks are then designed to estimate quantities (e.g., signal spectrum, correlation matrices) that are used to estimate these filters' parameters (Carbajal, Serizel, Vincent, & Humbert, 2020) (Hendriks & Gerkmann, 2011) (Heymann, Drude, & Haeb-Umbach, 2016) (Nugraha, Liutkus, & Vincent, 2016). These approaches are generally referred to as *hybrid* approaches, since they combine neural networks and signal processing-based filtering techniques.

The second set of approaches solely relies on neural networks to directly estimate the signals or the multichannel filters (Dowerah, Kulkarni, Serizel, & Jouvet, 2023) (Dowerah, Kulkarni, Serizel, & Jouvet, 2023) (Luo, Han, Mesgarani, Ceolini, & Liu, 2019) (Tolooshams, Giri, Song, Isik, & Krishnaswamy, 2020). In this latter case, the neural network parameters are fitted on a training set to optimize a cost function. Note that this differs from the first set of approaches, where the optimization problem is solved directly in an unsupervised fashion.

In this paper we study the behavior of three different methods. The motivation for choosing these is threefold. Firstly, their source code and trained parameters are available publicly. Secondly, these techniques can be applied in a binaural enhancement setup. Finally, they cover a wide variety of methodologies among the neural-based multichannel speech enhancement techniques. The first two methods integrate neural networks within signal processing-based filtering; thus, they belong to the first category mentioned above. One method is designed principally for compact microphone arrays and relies on a single-channel neural network, while the other is designed for distributed arrays and relies on a multichannel neural network. The last approach is fully based on neural networks; thus, it is representative of the second category mentioned above.

Even though these filters are applied in the time-frequency domain and depend on both the time frame $t$ and the frequency index $f$, these indices are omitted in the remainder of the paper when they are not necessary for a clarity purpose.

## Minimum variance distortionless beamformer

The target of the minimum variance distortionless beamformer (MVDR) is to minimize the noise contribution in the noisy mixture while the signal coming from the target direction (here, the target speech) is left unaltered (Capon, 1969). This can be expressed as finding a filter $\mathbf{W}$ as the solution to the following constrained optimization process:

$$\min_{\mathbf{W}} \|\mathbf{W}^H \mathbf{N}_i\|^2 \text{ subject to } \mathbf{W}^H \mathbf{d} = 1,$$

where $\mathbf{d} \in \mathbb{C}^M$ is the steering vector pointing towards the target source (note that in what follows we discard the indices $L, R$ for brevity). The constraint $\mathbf{W}^H \mathbf{d} = 1$ imposes that the signal coming from the target direction remains unaltered. This optimization problem is usually reformulated as follows:

$$\min_{\mathbf{W}} \mathbf{W}^H \mathbf{R}_N \mathbf{W},$$

where $R_N$ is the correlation matrix of the noise component. Solving this optimization problem leads to the so-called MVDR filter:

$$\mathbf{W}_{\text{MVDR}} = \frac{\mathbf{R}_N^{-1} \mathbf{d}}{\mathbf{d}^H \mathbf{R}_N^{-1} \mathbf{d}}.$$

To compute this filter, one must first estimate the steering vector $d$ and the noise correlation matrix $R_N$. The steering vector can be obtained from an explicit estimation of the direction of arrival of the target speech. A common alternative (which is notably used in this paper) is to estimate the steering vector as the principal component of the correlation matrix of the speech component $\mathbf{R}_S$. As a result, computing the MVDR filter relies solely on estimating the speech and noise correlation matrices.

Heymann et al. proposed to estimate these matrices using time-frequency masks computed with a recurrent neural network (Heymann, Drude, & Haeb-Umbach, 2016). The noisy mixture is fed as input to the neural networks, which returns a speech mask coefficient $M_S$ indicating the amount of speech that is present in the mixture in each time-frequency point, as illustrated in Figure 3. The correlation matrix of the speech component is then obtained as follows:

$$\mathbf{R}_S(f) = \frac{1}{T} \sum_{t=1}^{T} M_S(t,f)\mathbf{X}(t,f)^H \mathbf{X}(t,f).$$

( 1 )

A similar process is conducted to obtain the noise correlation component.

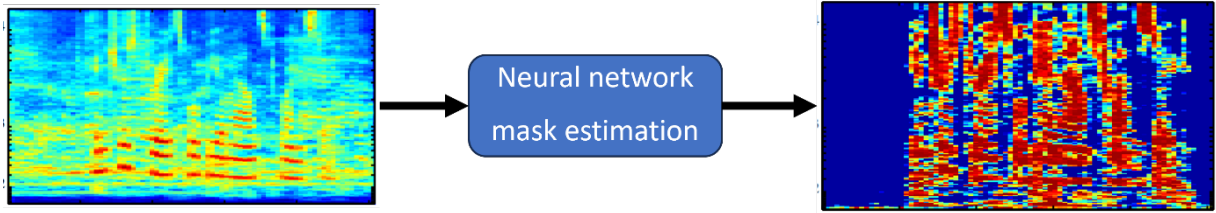

*Figure 3 - Neural network-based mask estimation. The neural network is fed noisy signal at the input and provides a mask that indicates the amount of speech in each time-frequency bin.*

Note that in practice in our setup, we compute a filter for each hearing aid, thus using either $X_R$ or $X_L$ instead of $X$ in the equation ( 1 ).

### Distributed multichannel Wiener Filter

The goal of the multichannel Wiener Filter (MWF) is to estimate the speech component at an arbitrary reference microphone $S_{\text{ref}}$ (Doclo & Moonen, 2002). This can be formulated as the minimization of a mean squared error criterion:

$$\min_{\mathbf{W}} \|\mathbf{W}^H \mathbf{X} - S_{\text{ref}}\|^2,$$

where $S_{ref} = d_{ref}^H X$, and $\mathbf{d}_{ref}$ is a vector whose entries are equal to 0, except for the one corresponding to the reference channel where it is equal to 1. Solving this optimization problem leads to the MWF formula:

$$\mathbf{W}_{MWF} = \mathbf{R}_X^{-1} \mathbf{R}_S \mathbf{d}_{ref}.$$

The computation of the MWF then relies only on the correlation matrices $\mathbf{R}_X$ and $\mathbf{R}_S$, that can be estimated with neural networks similarly as for the MVDR.

Variants of the MWF have been proposed. The speech distortion weighted MWF allows for adjusting the trade-off between reducing the noise and minimizing the speech distortion (Spriet, Moonen, & Wouters, 2004). The so-called generalized eigen value decomposition MWF (GEVD-MWF) selects the reference channel in the generalized eigenvalue space, allowing for a more robust filter, in particular in very noisy conditions (Serizel, Moonen, Van Dijk, & Wouters, 2014).

Here we consider binaural speech enhancement as a particular case of distributed ad-hoc antennas. Bertrand et al. proposed a variant of MWF that is adapted to this case (Bertrand & Moonen, 2010). This latter algorithm was originally proposed assuming a perfect voice activity detection to estimate the correlation matrices. It was later adapted to replace the voice activity detection by a two-step neural network-based mask estimation method called Tango (Furnon, Serizel, Essid, & Illina, 2021). In the first stage the mask is obtained for one local channel, similarly to what we described for MVDR. In the second stage, the neural network has access to the signal recorded at the ear of interest as well as the signals filtered at the contralateral ear in the first step. These signals are used jointly to estimate the masks (Figure 4). In the paper we will use this latter approach where the filtering is part is a GEVD-MWF.

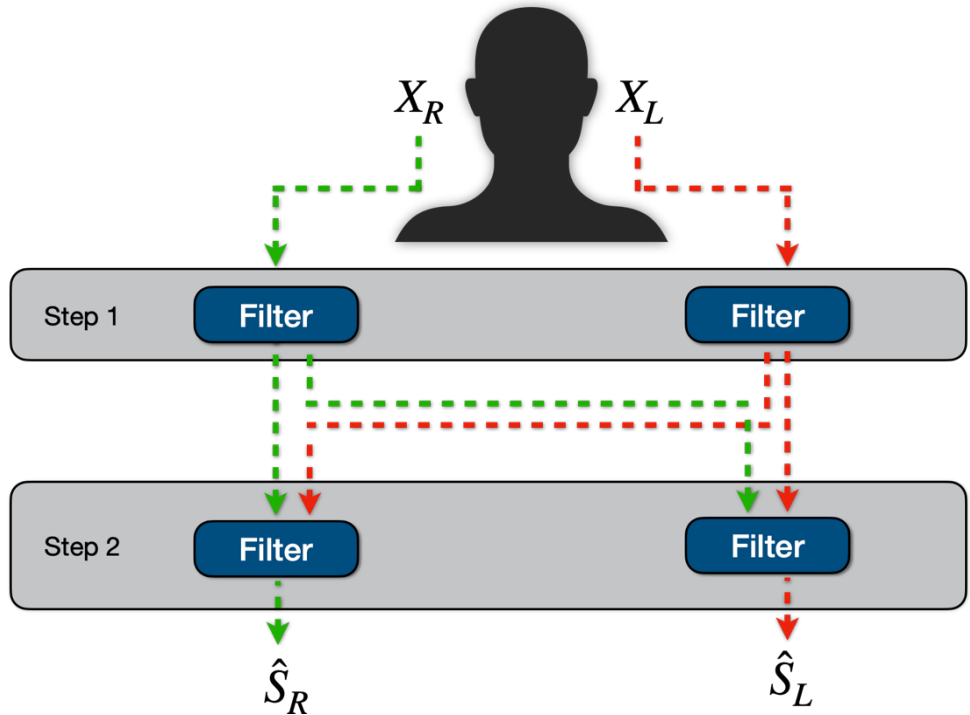

*Figure 4 - Neural network based distributed MWF in the binaural case.*

### Filter-and-sum network beamformer

The last beamformer that we study in this paper is a beamformer that relies mainly on neural networks, the so-called Filter-and-sum Network (FaSNet) beamformer (Luo, Han, Mesgarani, Ceolini, & Liu, 2019). More precisely, the beamformer itself is computed with neural networks and directly applied to the signals recorded by the microphones.

This method also operates in two stages. In the first step, a beamformer is computed to extract a filtered speech signal at an arbitrary reference channel. In the second step, this filtered reference signal is used to compute pairwise beamformers for all other channels. Since the beamformer coefficients are computed with a neural network, there is possibly more flexibility regarding the target cost function. Unlike the previous hybrid approaches, the cost function does not need to be convex to allow for a closed form solution as it optimized numerically on the training set. In the paper, the FaSNet beamformer is trained to optimize a scale-invariant signal to distortion ratio (SI-SDR) (Le Roux, Wisdom, Erdogan, & Hershey, 2019).

## Methodology

This section details the methodology we adopted in our study. First, we present the pipeline used for simulating multichannel speech data in realistic noisy conditions. Then, we introduce the evaluation metrics. Finally, we describe the phoneme categories and the classification method that allows for a fine-grain assessment.

## Data generation

A fully realistic benchmark would involve conducting evaluations on real signals that contain speech and noise with respect to the desired configurations (e.g., regarding the position, amount, and type of noise). Yet, acquiring such a diversity of signals is complex, costly, and time intensive. Relying solely on real-life signals constrains the evaluation to a limited set of scenarios. Thus, simulations emerge as a viable alternative to easily generate a large quantity of signals in various scenarios, closely resembling real-world settings. To that end, we simulate mixtures that replicate diverse acoustic scenarios with target speech contaminated by interfering noise. This process relies on the availability of a clean speech corpus, a noise corpus, and functions that describe room acoustic properties.

### Speech data

The speech data used to simulate mixtures comprises 1000 speech signals extracted from the test set of Librispeech (Panayotov, Chen, Povey, & Khudanpur, 2015). This dataset is a comprehensive corpus encompassing approximately 1000 hours of English speech. The data is obtained from audiobooks within the LibriVox project, where audio recordings have been aligned with their corresponding texts and partitioned into short segments. The signals are sampled at 16 kHz.

### Noise types

We consider both synthetic and recorded noise types (see Figure 5). This approach allows us to observe and understand the different behavior of the speech enhancement algorithms described above.

First, we consider white noise, which is characterized by its uniform frequency distribution. As outlined in Keith and Talis (Keith & Talis, 1972), white noise stands out as a standard reference in hearing loss audiological testing.

We also consider speech-shaped noise, a type of noise signal designed to mimic the average frequency characteristics of natural human speech. Unlike white noise, which has a uniform frequency distribution, speech-shaped noise is crafted to simulate the spectral characteristics of spoken language. This type of noise replicates the statistical properties of speech sounds, including the energy distribution across frequencies. The goal is to create a background noise that closely resembles the tonal features of real-life speech: this provides a more ecologically valid testing environment than white noise but allows for a more controlled setup than real-life noise.

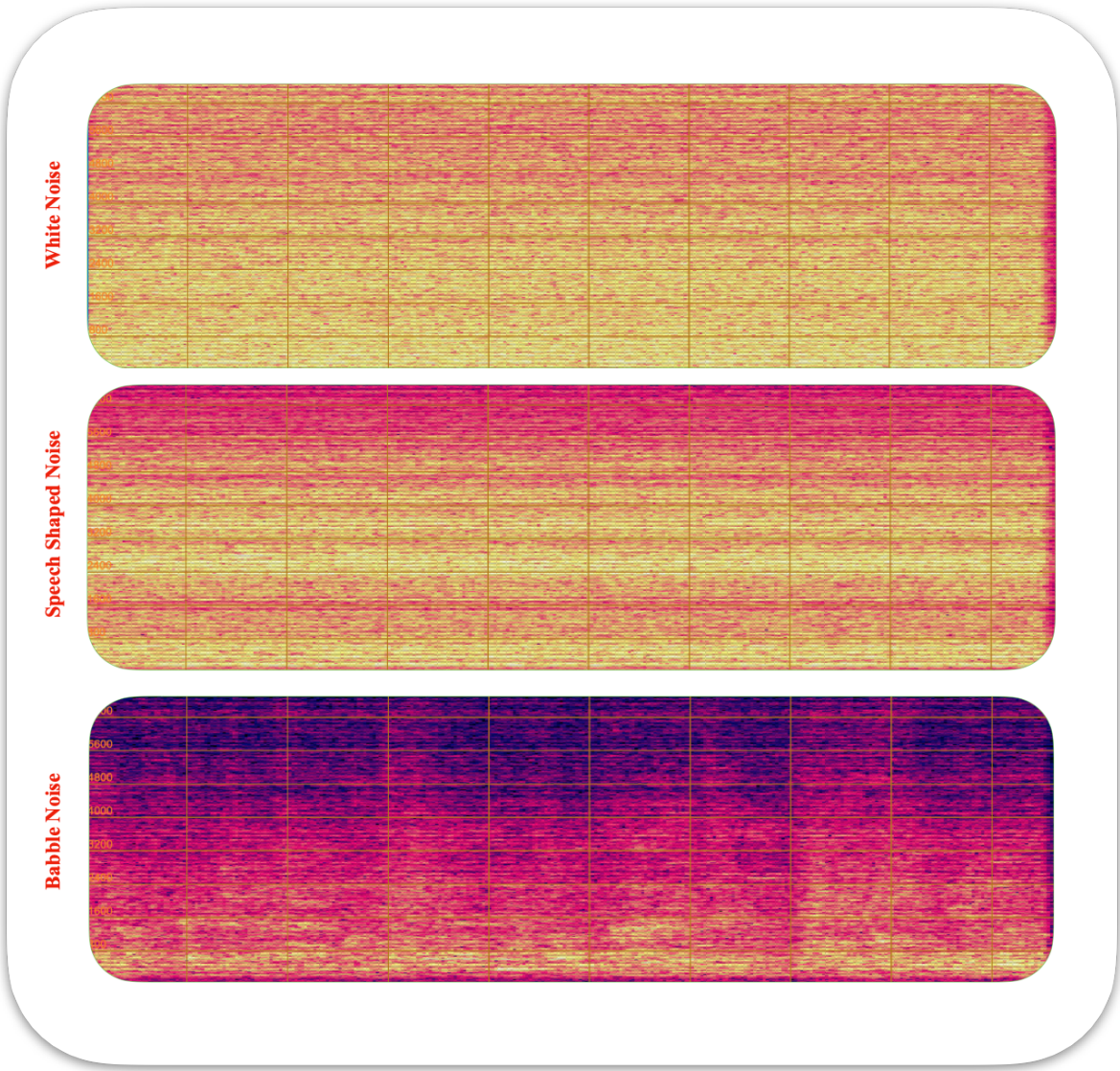

*Figure 5 - Spectrograms of white noise, speech shaped noise and babble noise.*

To generate speech-shaped noise, we use a speech signal from Librispeech that does not belong to the subset we used for generating mixtures. First, we compute its discrete Fourier Transform (DFT) in order to transform the signal into the frequency domain. We keep the signal's DFT magnitude, and we randomize its phase, then we revert it back to time-domain via inverse DFT. This produces a noise signal that mirrors the spectral properties of the speech signal.

Finally, we consider a babble noise signal taken from Freesound (Font, Roma, & Serra, 2013). The selected clip was recorded in a restaurant during a lunch break. This audio signal consists of a soundscape where multiple people are conversing simultaneously. Unlike synthetic noise signals (e.g., white noise or speech-shaped noise), babble noise comprises overlapping speech from various speakers in the same acoustic scene, which introduces a higher level of acoustic complexity. This complexity mirrors real-world scenarios to investigate how individuals with hearing loss navigate challenging auditory scenes.

## Room Impulse Responses

A room impulse response (RIR) is a filter that describes the impact of sound propagation within a room from the position where the sound source is emitted to the microphone where it is recorded. It encapsulates information about both the relative positions of the source and microphone inside the room, and intrinsic characteristics of the room, such as its size, geometry, and absorption parameters. Mathematically, the sound at the microphone can be described as the convolution between the RIR and the emitted source signal.

RIRs can be obtained by either measurement or simulation. RIR simulation is based on algorithms that generate an RIR via a room acoustic model, thus it does not require to conduct any physical measurement. On the other hand, a measured RIR is specific to the room and spatial configuration where the measurement was conducted. This process allows for a more accurate representation of the acoustic environment, even though it is more limited in terms of diversity of acoustic scenarios. Here we will use this latter approach to generate the signals used to evaluate the speech enhancement algorithms.

One method to measure an RIR, as proposed by Novak & al. (Novak, Lotton, & Simon, 2015), involves playing a reference audio signal at the location of a given source (e.g., speech or noise), and recording the reverberated signal at the position where the mixture will be created, typically where the listener stands (or more precisely at the listener's left or right ear).

As a reference signal, we use a sweep signal, which has a constant amplitude and whose frequency varies linearly over time from 20Hz to 20kHz. To capture the reverberated sweep signal, a Portable Hearing Laboratory (PHL) equipment has been set up on a KEMAR head and torso model. The PHL, serving as a hearing aid simulator, incorporates a configuration with four microphones (two for each ear), facilitating the capture of four distinct speech channels, as commonly found on hearing aid devices. The KEMAR, which is designed to mimic human-like head and torso features, is employed to simulate the head's orientation in real-world scenarios. The sweep signals are played at three distinct angles: 0° (front), 45°, and 90°, where the angles are defined according to the orientation of the head (see Figure 6). This allows us to simulate a setup where the speech signal is placed in front of the listener (using the RIR at 0°), and where the noise source can be placed laterally at either 45° or 90°.

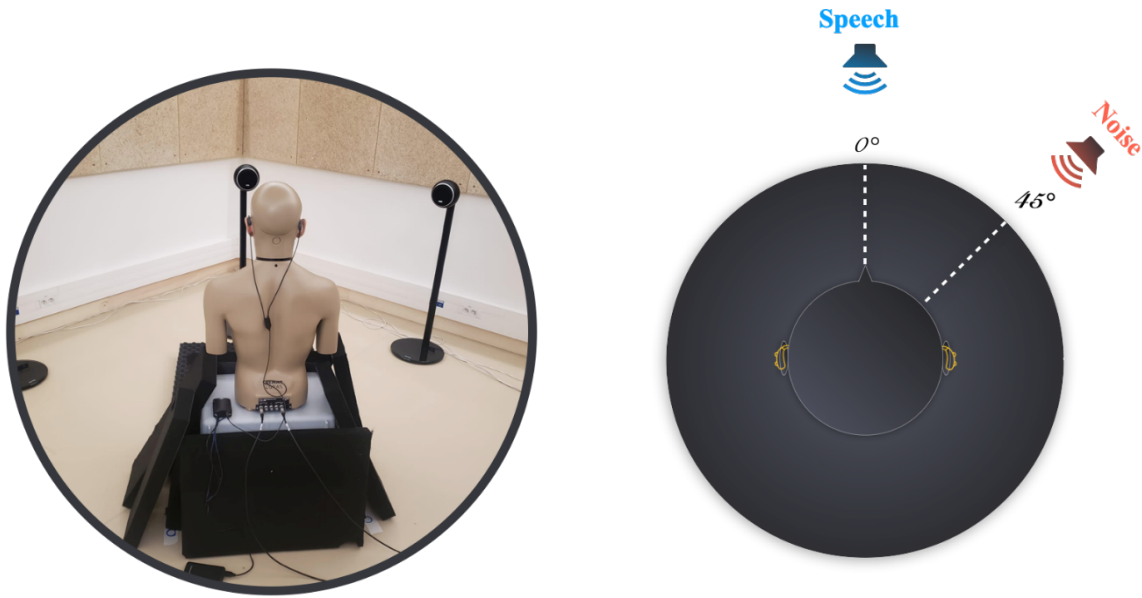

*Figure 6 - Acoustic simulation setup using the KEMAR and PHL devices (left), and spatial configuration of the speech and noise sources (right).*

## Mixtures

In order to build noisy mixtures, we first need to adjust the relative speech and noise levels. To that end, we apply a gain factor to the noise source, and we control the amount of noise via the signal-to-noise ratio (SNR). In its simplest form, the SNR is defined as:

$$SNR = 10.\log10\left(\frac{\|s\|^2}{\|n\|^2}\right),$$

and expressed in decibels (dB), where $s$ and $n$ respectively denote the anechoic speech and noise signals. In practice, the SNR is computed by only considering segments where the speech signal is active. Note that the SNR is adjusted by considering source signals (before applying the RIR). Subsequently, the clean speech and the scaled noise are convolved with the RIRs as detailed in the previous section, resulting in a mixed audio signal by summing the two convolved signals. This process effectively simulates the acoustic characteristics of a real-world environment.

## Evaluation metrics

In our evaluation, we use objective metrics to compare the target speech with the speech estimated with the different speech enhancement algorithms at utterance and phoneme levels. Traditionally, speech enhancement algorithms are tested using metrics that measure the quantity of interference, artifacts and distortions that remain in the estimated speech. To that end, we use the BSS eval metrics from Vincent et al. (Vincent, Gribonval, & Févotte, Performance measurement in blind audio source separation, 2006), originally tailored for source separation applications, but widely used for speech enhancement. Let us consider the following decomposition of the error between the target speech signal $s_{\text{target}}$ and its estimate $\hat{s}$ at a reference microphone:

$$\hat{s} = s_{\text{target}} + e_{\text{interf}} + e_{\text{artif}},$$

Where $e_{\text{interf}}$ and $e_{\text{artif}}$ denote the interference and artifacts errors, that is, the contributions of the non-target source(s). Here the interference is the contribution of the noise source, and the artefacts represent other types of distortions (e.g., burbling noise) in the speech estimate. In what follows, we remove the indices $i, m$ for brevity. From this decomposition, we define the following signal-to-distortion ratio (SDR), signal-to-artifacts ratio (SAR), and signal-to–interference ratio (SIR) as follows:

$$SDR = 10.\log_{10} \frac{\left\|s_{\text{target}}\right\|^2}{\left\|e_{\text{interf}} + e_{\text{artif}}\right\|^2},$$

$$SAR = 10.\log_{10} \frac{\left\|s_{\text{target}} + e_{\text{interf}}\right\|^2}{\left\|e_{\text{artif}}\right\|^2},$$

$$SIR = 10.\log_{10} \frac{\left\|s_{\text{target}}\right\|^2}{\left\|e_{\text{interf}}\right\|^2}.$$

These ratios are expressed in dB, and for all metrics, higher is better. Since they are computed using estimated speech signals, that is, the *outputs* of the speech enhancement algorithms, we will refer to them as SDR_out, SAR_out, and SIR_out.

Note that in order to quantify the actual noise reduction achieved by the speech enhancement algorithm, it is necessary to compare a given output metric to a reference initial value. To that end, we calculate the metrics by replacing the estimated speech with the noisy mixture: since the resulting metrics are computed at the *input* of the algorithms (before any processing), we refer to them as SDR_in, SAR_in, and SIR_in.

In particular, SIR_in measures the ratio of desired speech to background noise as it is received at the ear level. In our setup, since we do not consider any additional measurement noise (e.g., induced by the recording device), this is equivalent to a signal-to-noise ratio, except it is calculated using the images instead of the source signals. As such, this metric is critical for understanding the impact of a room's acoustics at the listener's ear and serves as a reference point for the mixture signal quality (i.e., before enhancement). By comparing SIR_out and SIR_in, we can quantify the actual noise reduction achieved by the speech enhancement model.

Note that in theory, there are no artefacts in the input signals, so SAR_in is infinite and the SDR_in is equal to SIR_in. Therefore, we will not consider these metrics when presenting our results.

When evaluating speech enhancement algorithms, these metrics are typically computed at the utterance level and aggregated over several sentences to obtain a consolidated metric. This process overlooks the potential performance variability of algorithms depending on the phonetic content of the speech signals.

## Phoneme classes

In our study, we investigate the evaluation of speech enhancement algorithms at the phoneme level. In this regard, we perform phoneme segmentation of clean speech signals. This process involves using a phoneme recognizer to estimate the boundaries of each phoneme within a speech signal.

We used the Montreal Forced Aligner (MFA) (McAuliffe, Socolof, Mihuc, Wagner, & Sonderegger, 2017) to align spoken audio recordings with their corresponding phonetic transcriptions. In our analysis, we utilized the "English MFA dictionary v2.2.1" version, which has been trained on a dataset comprising 95,278 words. This ensured the alignment model's proficiency in handling a diverse range of phonetic patterns and nuances present in the clean speech data.

As shown in Figure 7, the speech dataset comprises 55 phonemes, each with varying frequencies. Most frequent phonemes include /ə/ with 2373 occurrences, /ɪ/ with 2022, and /n/ with 1972. The least frequent phonemes occur less than 10 times in the whole dataset. Hence a study at the phoneme level would hardly lead to any statistically significant outcome on these phonemes.

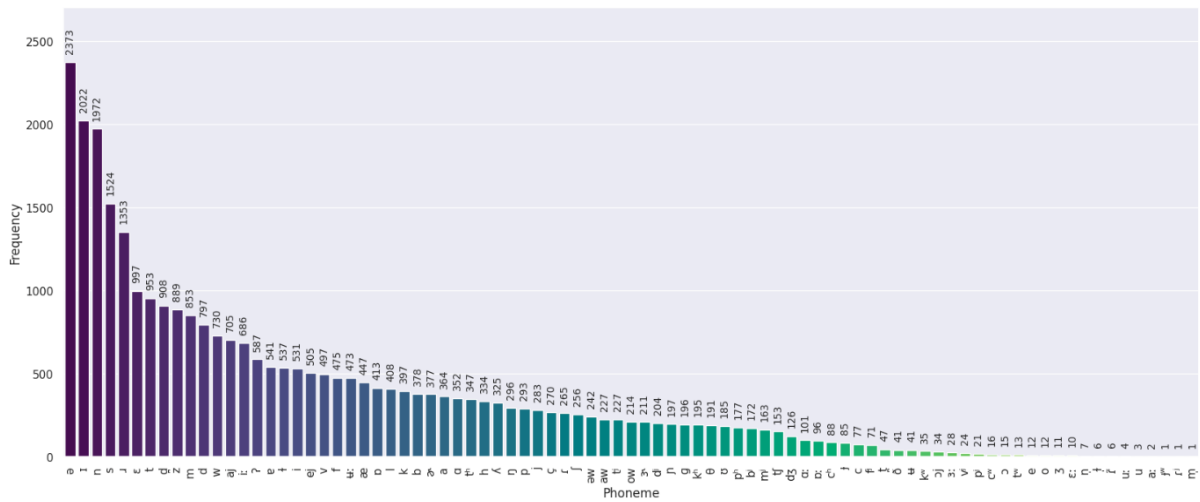

*Figure 7 - Phoneme distribution in our speech dataset*

To simplify the analysis, we group phonemes into categories. This classification relies on a slightly modified version of the MFA IPA chart, as we included an additional vowel class for the near-close near-front unrounded phoneme /ɪ/, and the near-close near-back rounded phoneme /ʊ/. This decision was driven by the important presence of the unrounded phoneme within the dataset, prompting us to investigate the near-close performance and behavior in the evaluations. Moreover, we propose to include both /e/ and /ej/ in the close-mid category as the number of /o/ and /ow/ is very low.

As illustrated in Figure 8, the most prominent phoneme categories in our dataset plosives, open-mid, and nasals, occurring 6214, 4287, and 2489 times, respectively. The close-mid, affricate, and tap phonemes categories add to the phonetic diversity, albeit as the least frequent, enriching the overall representation of speech.

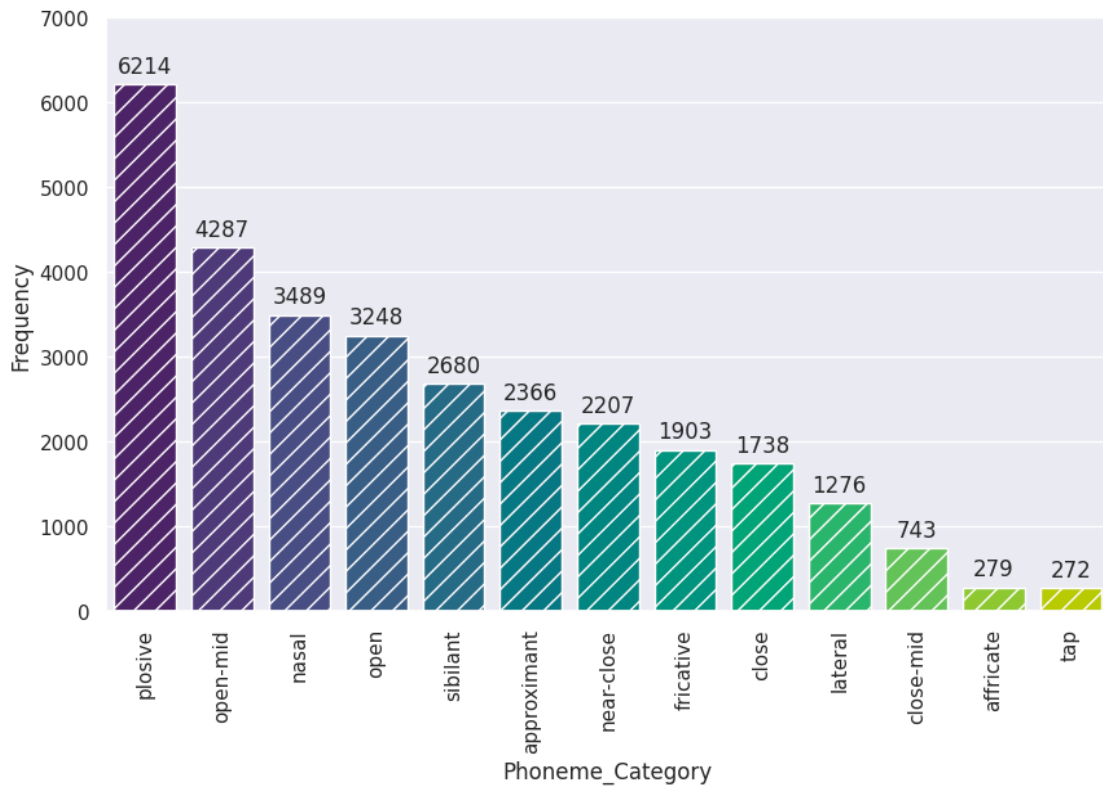

*Figure 8 - Distribution of phoneme per categories in our speech dataset*

### Experimental setup

In the experiments conducted in this paper we used the measured RIRs from the BinauRec dataset of Delebecque and Serizel. (Delebecque & Serizel, 2023),which complies with the protocol outlined in Section Room Impulse Responses. The RIRs in Binaurec are recorded in a single room and correspond to sources positioned at 0°, 45° and 90°. These RIR were convolved with speech and noise signals to create the 4 channels audio mixtures that serve as the input for the speech enhancement models. Noisy mixtures are built using an SNR of -5, 0, or 5 dB. For each ear, the front microphone of the hearing aid device is selected as the reference microphone.

As outlined in Section Overview of Multichannel Speech Enhancement, we selected three speech enhancement algorithms whose pre-trained weights are available online. All the models are used in their default setup. The MVDR and FasNet models are implemented in the ESPnet toolbox (Li, et al., 2021), and the corresponding weights can be readily downloaded from the toolbox. Both models have been trained on the CHiME-4 dataset (Vincent, Watanabe, Nugraha, Barker, & Marxer, 2017). The Tango model (Furnon, Serizel, Essid, & Illina, 2021) is trained on the same dataset as in the original paper and its code and pre-trained weights are available online.

In our study, we compute the BSS eval metrics using the mir_eval library (Raffel, et al., 2014). For a reproducibility purpose, our code is available online at: https://github.com/Nasseredd/SE-Ph-Eval/.

# Experimental Results

This section presents our experimental results. First, we detail the results at the utterance level, as speech enhancement evaluations are typically conducted. Then, we delve into a finer-grain evaluation at the phoneme level.

## Evaluation at the utterance level

### Comparison between the left and right microphones

This first experiment compares the results obtained for the two reference microphones (left and right). We operate in an asymmetrical scenario, since the noise source is placed on the right side of the recording device in our setup (see Figure 6). This investigation aims to clarify how the noise's spatial orientation affects the binaural algorithms' performance on each microphone.

Figure 9 displays the results of the algorithms' effectiveness to enhance the speech at each ear (note that these results are averaged across noise types and positions, SNR levels, and models). In line with our expectations, the influence of noise is more pronounced on the right ear. The right ear exhibits lower input SIR, which can be attributed to the proximity of the right reference microphone to the noise source. On the contrary, the head shadow effect impacts the sound propagation to the left ear and the input SIR is larger.

Even though assessments of the estimated speech (output) indicate a better absolute performance on the left microphone, it is noteworthy that the relative improvement is more important for the right microphone. Specifically, we observe that the SIR improvement is more substantial on the right side (10,60 dB) than on the left side (2,86 dB). This outlines that the right-side microphone benefits more from the binaural property of the speech enhancement algorithm. Nonetheless, from an audiology standpoint it is more important to improve the sound quality at the best ear (here: the left) (Bronkhorst & Plomp, 1988), thus in what follows we will focus our analysis on the results for the left reference microphone.

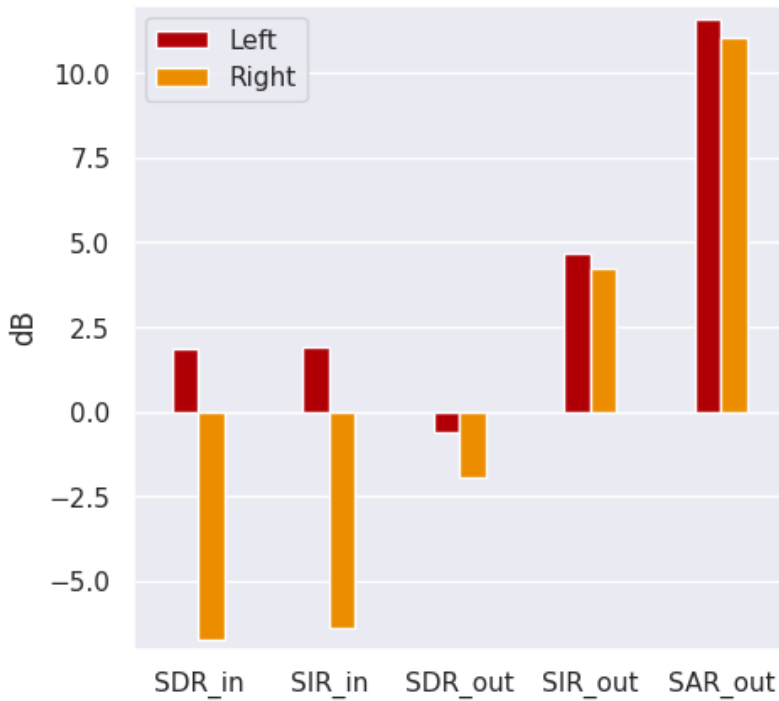

*Figure 9 - Comparison between the left and right microphones.*

### Comparison between noise types

In this experiment we investigate the influence of the noise type on performance. We consider three noise types: white noise, speech shaped noise and babble noise. The results are averaged across SNR levels and models and displayed in Figure 10.

First, we observe an overall consistent performance for the white noise and the speech shaped noise. The babble noise is more challenging for models to deal with than the other noise types, as indicated by the corresponding low values of input SDR, SIR and SAR. As the focus of the paper is not to analyze the performance of speech enhancement algorithms under challenging scenarios, but rather to understand their behavior at a fine-grained scale we will focus on white noise and speech-shaped noise. These two noise types are indeed easily controlled, yet they allow for investigating different masking effects on the phonemes. Note also that the speech-shaped noise can serve as a proxy to the babble noise.

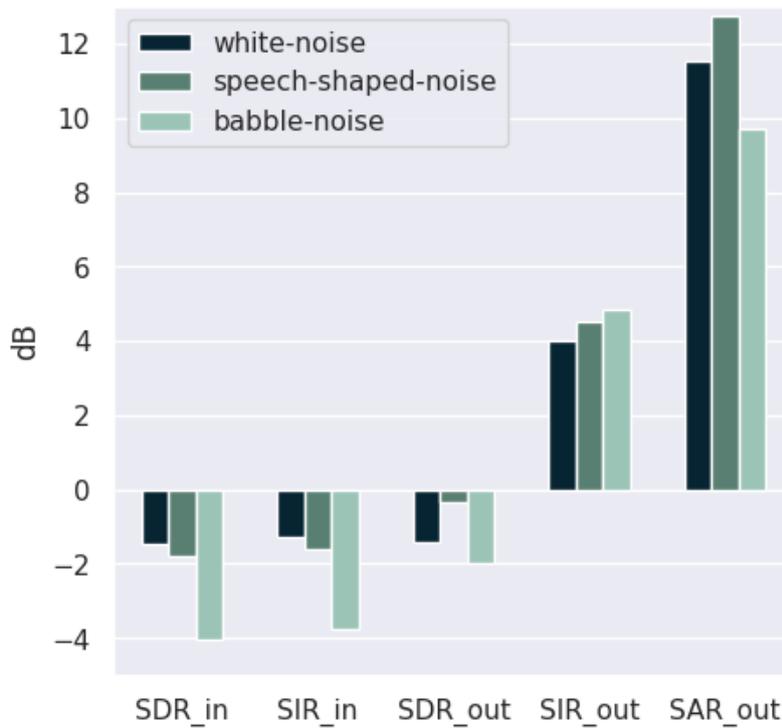

*Figure 10 - Comparison between the white noise, speech-shaped noise and babble noise.*

### Impact of the noise location

This experiment analyzes the impact of the noise location on the performance of the speech enhancement algorithms. The noise source can be positioned at either 45° or 90° of the listener (see Figure 6). The results are averaged across noise types, SNR levels, and models, and presented in Figure 11.

First, we remark that the input SIR is slightly higher when the noise is placed at 90°. This was expected since these results correspond to the left-side microphone, which is less contaminated with noise when the source is placed on the opposite side of the head rather than at 45°. Likewise, we also observe that speech enhancement algorithms exhibit a higher performance when the noise is placed at 90° compared to 45°, as indicated by the output SDR and SIR. Nevertheless, the differences between the two scenarios are not very important in terms of both input and output metrics. Therefore, selecting one of the two scenarios would not impact the overall analysis greatly. In the rest of our study, we will focus on the 45° scenario that is potentially more challenging for the speech enhancement algorithms.

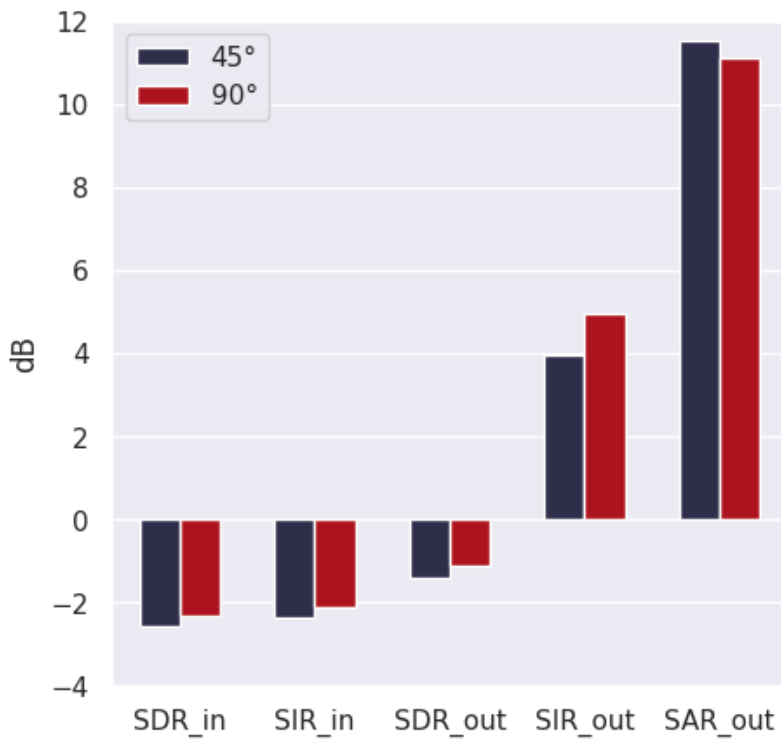

*Figure 11 - Comparison between noise angles at 45° and at 90°.*

Impact of the SNR level

Figure 12 presents the impact of varying the mixture SNR levels on both the input and the output evaluation metrics.

First, comparing the SNR (computed using the anechoic sources) and the input SIR (computed using the image sources) highlights the influence of room acoustics on the mixtures at the ears' level. On average, this phenomenon results in a drop of approximately 2 dB in terms of noise level for the three scenarios. This is justified by the propagation in the room and the position of the sources with respect to the walls (the noise source is located close to a corner in this case which tends to amplify the sound transmitted). This underscores the importance of computing the SIR at the microphone as a reference value and not relying on the SNR that is set on the dry sources.

Overall, we observe that the algorithms' performance improves as the mixture SNR increases. We can also observe an important drop in all the metrics for the scenarios where the SNR is -5 dB. This indicates that it will probably be useful to control the performance of the algorithms at different input SNR. Yet, for the simplicity of the analysis, in our phoneme-level experiments, we will focus mainly on the scenario with a 0 dB SNR. Indeed, this setting is sufficiently challenging for the task at hand and allows us to examine the relative performance of speech enhancement algorithms without inducing an excessive degradation of the signal. Still, we will examine the impact of the mixture SNR on specific phonemes, where it might deserve some finer-grain analysis.

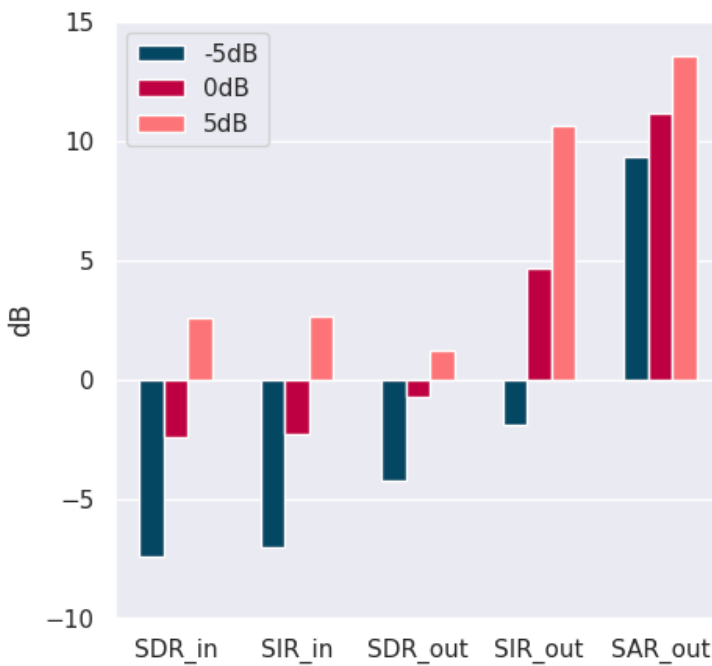

*Figure 12 - Comparison between SNR levels (-5dB, 0dB, 5dB).*

## Evaluation at the phoneme level

We now delve into our evaluation at the phoneme level, which we first illustrate by the following introductive example. Figure 13 displays the spectrogram of a clean speech signal, as well as the spectrograms of this same signal contaminated with noise, and where some specific phonemes are highlighted.

The spectrograms display for instance the phonemes 'g' and 's' across different noise conditions. In the clean speech, the phoneme 'g' shows distinct vertical striations representing its voiced nature with a rich harmonic structure. The phoneme 's' is characterized by a high-frequency, almost texture-like pattern, indicating its sibilant, unvoiced nature.

When white noise is added to the clean speech, we observe that the low frequency harmonics of the phoneme 'g' remains relatively intact. On the other hand, the sibilant 's' is strongly affected by this noise, since its energy is more concentrated in the higher frequencies where the white noise also contains energy.

With speech-shaped noise, the impact on the phoneme 'g' is less uniform. The noise follows the speech spectrum, filling in the silent gaps and making it difficult to discern the phoneme's harmonic patterns. The phoneme 's' remains relatively discernible, but its crisp edges are somewhat softened, and the definition between silence and sibilance is less clear.

The presence of babble noise introduces a more complex interference. The phoneme 'g' is disrupted by the varying intensities and frequencies of overlapping speech, obfuscating its harmonic structure. Conversely, while still visible due to its high-frequency content, the phoneme 's' competes with similar sounds from the babble, which can make it challenging to isolate from the background chatter.

This analysis underscores the critical importance of evaluating speech enhancement algorithms at the phoneme level. The differences in how various phonemes are affected by different types of noise highlight the nuanced challenges faced by the speech enhancement systems. Voiced phonemes, with their rich harmonic structures, and unvoiced phonemes, with their high-frequency energy, require different enhancement strategies to overcome the masking effects of noise. Understanding these varied impacts is essential for designing speech enhancement algorithms that can effectively disentangle and clarify the important elements of speech, ensuring that each phoneme, regardless of its unique acoustic properties, is accurately reproduced and easily discernible, even is adverse listening conditions.

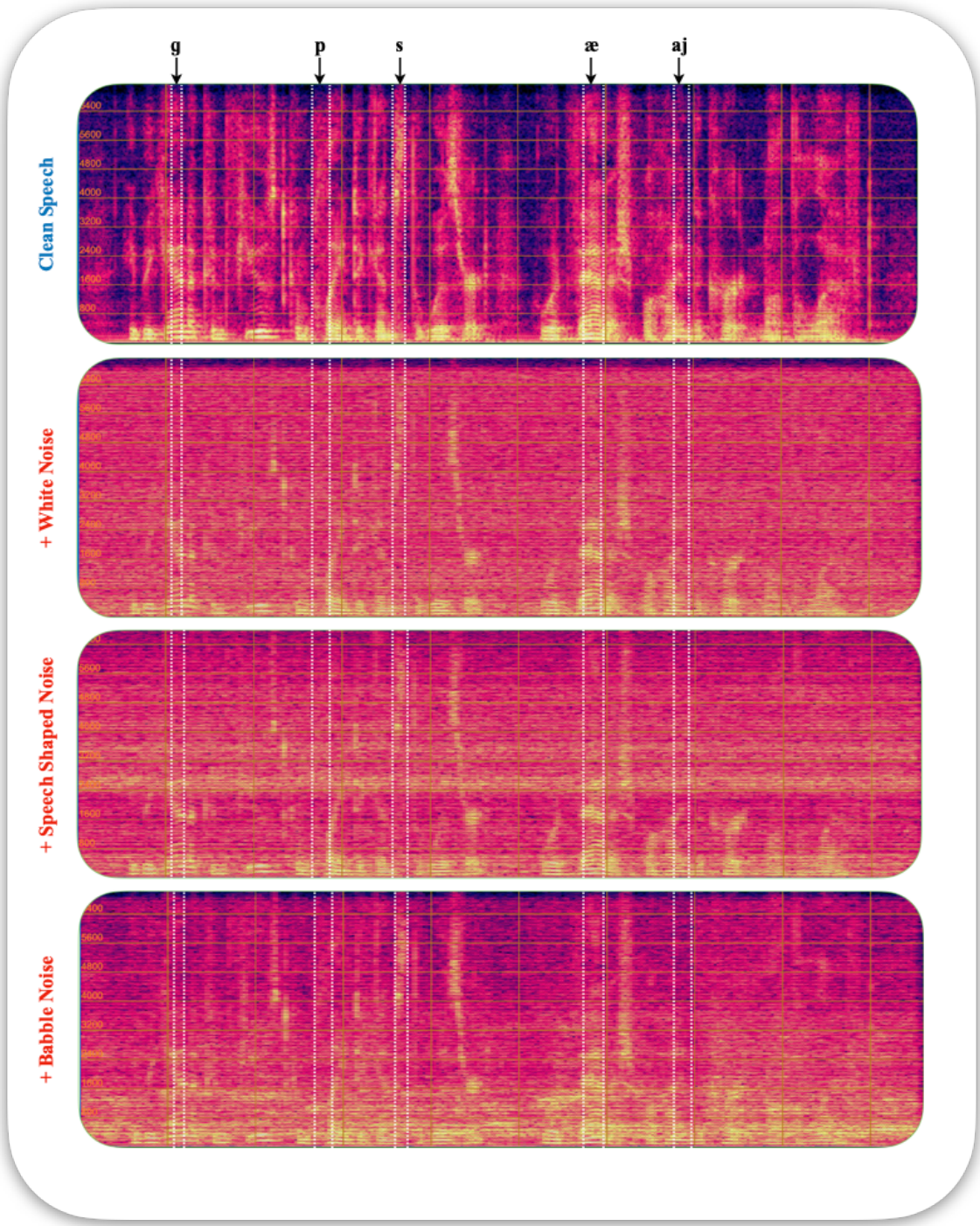

*Figure 13 - Spectrogram of clean speech and mixtures with different types of noise on the utterance "He began a confused complaint against the wizard who had vanished behind the curtain on the left" segmented into phonemes.*

## An overview of the results at the phoneme level

We first present an overview of the results at the phoneme level. In all the following experiments (except for the last one), the mixtures are at 0 dB SNR. The results are displayed in Figure 14. A notable observation is that plosives, fricatives and taps are the most impacted by the noise. These are also the phoneme categories on which the speech enhancement algorithms perform the worst.

The experiment reveals that, on average, the speech enhancement models yield a large improvement in all the other phoneme categories. They particularly reduce the amount of noise while introducing only a controlled amount of distortion and artifacts. However, this positive trend in phoneme categorization contrasts with the findings at the utterance level. Indeed, the evaluation of artifacts at the utterance level tends to overestimate the performance across all phoneme categories. This suggests that while the models perform well at refining speech at the phoneme level, their effectiveness may be overstated when considering the broader context of complete utterances.

Overall, we observe different trends per phoneme categories, which motivates us to analyze these results in more depth. We conduct such an analysis in the following experiments, for which we select specific categories of phonemes such that the comparison is made clearer.

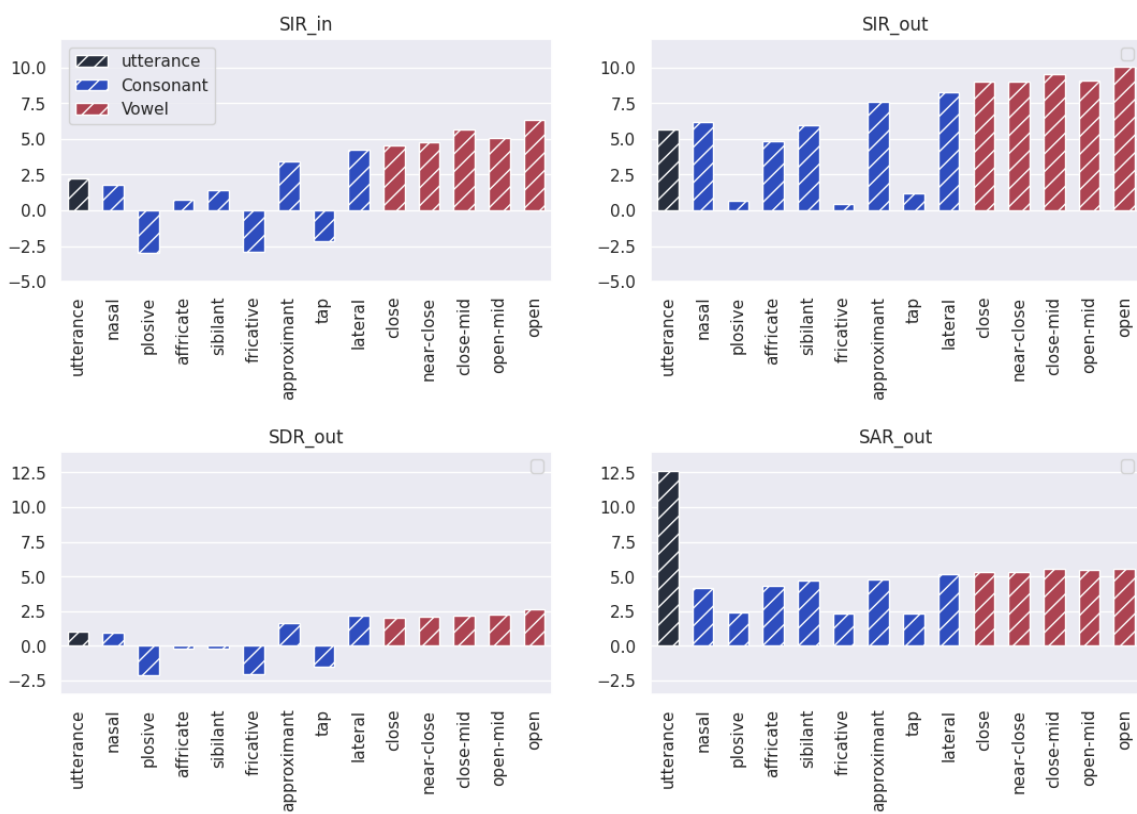

Figure 14 - Evaluation results across phoneme categories (the results at the utterance level are also reported). Results are averaged over algorithms.

### Impact of the noise type on plosive, approximant and open phonemes

In this experiment, we analyze the results (displayed in Figure 15) for opens, approximant, and plosive phonemes with respect to the noise type (white noise or speech-shaped noise). In examining the outcomes across various metrics, it is apparent that the general trend persists regardless of the noise type. Nonetheless, there is a slightly higher SIR improvement when speech-shaped noise is present as opposed to white noise. This can be explained by the fact that speech-shaped noise is commonly used for training speech enhancement models. It is interesting however to see that the performance in terms of SAR remains consistent across phoneme categories, regardless of the noise type. This indicates that the SIR improvement does not occur at the costs of a lower SAR, while such a trade-off is usually observed in speech enhancement algorithms.

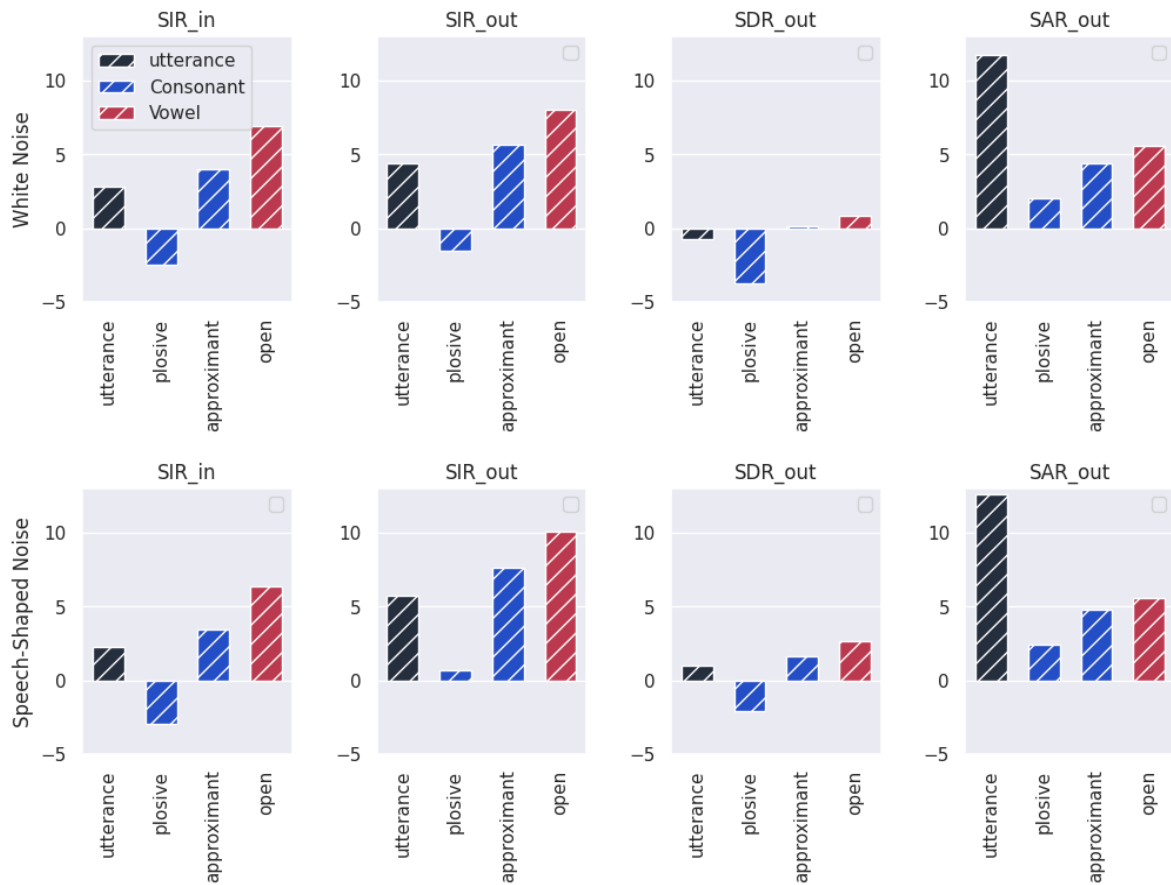

*Figure 15 – Evaluation results per noise types on plosive, approximant and open phonemes. Results are averaged over algorithms.*

## Comparison of the algorithms on nasals, affricates, and sibilants

We now conduct a comparative analysis of the three speech enhancement algorithms on the phoneme categories of nasals, affricates, and sibilants. The results are presented in Figure 16. The input SIR indicates that the nasals are the least degraded by the noise (with the white noise having slightly less impact than the speech-shaped noise).

As for the residual interference and distortions in the estimated speech, the performance varies with the noise type. Tango outperforms other models in mitigating interference and distortions with the presence of white noise. In contrast, when using speech-shaped noise, MVDR appears to be the best at reducing interference and artifacts, whereas Tango is superior for reducing distortions. FasNet appears to deteriorate the speech signal when white noise is involved, inadvertently increasing the interference beyond the original input level. This could potentially be due to a mismatch with the training conditions. Nevertheless, FasNet notably improves the SIR in the presence of speech-shaped noise, especially for nasal phonemes, indicating its effectiveness in enhancing certain aspects of speech.

Across different noise conditions, Tango exhibits robustness, consistently improving the SIR, more so for sibilants than for nasals and affricates. This suggests that Tango presents a balanced performance across various acoustic noise scenarios on the three phonemes categories. MVDR performs well on nasals regardless of the noise type but its performance on affricates and sibilants is always lower than for nasals.

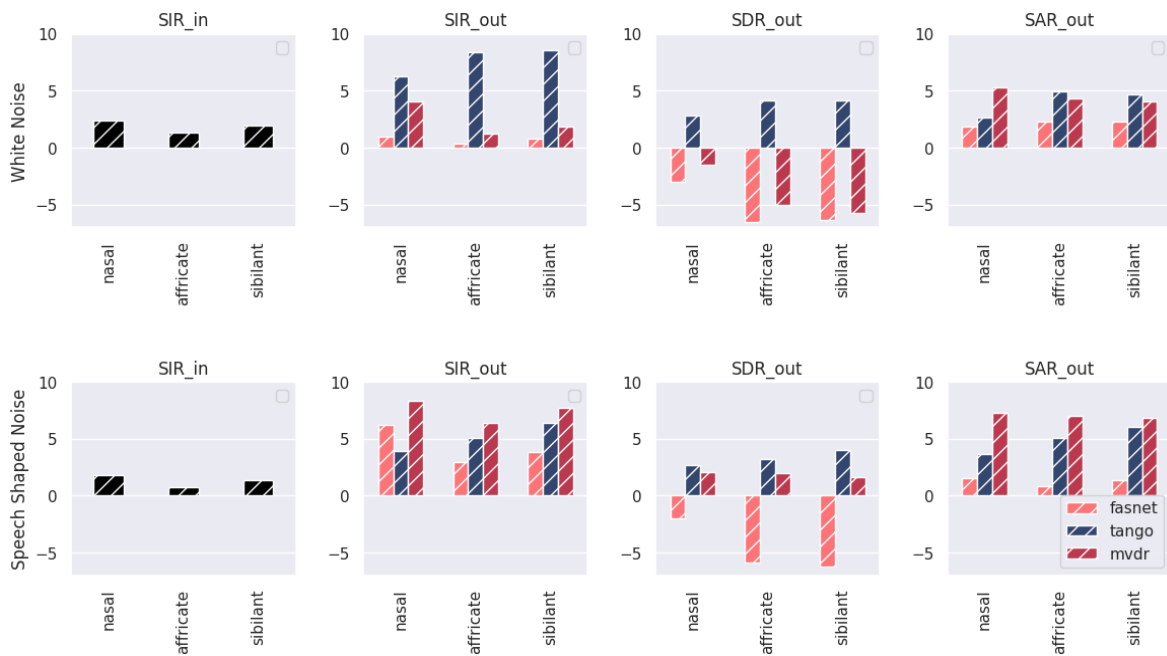

*Figure 16 – Performance of the algorithms on nasals and sibilants on mixtures.*

## Comparison of the algorithms on vowels

We now compare the three algorithms on vowels, which are organized into the close, near-close, close-mid, open-mid, and open phoneme categories. The results are presented in Figure 17.

In terms of SIR_in, the initial conditions for all vowel categories are favorable under both noise types, indicating that the vowels are less affected by noise than the consonant at the input stage. Besides, the input SIR remains consistent across all vowel categories.

The analysis of the speech enhancement algorithms reveals that Tango is the most effective model for minimizing distortions, performing with both white noise and speech-shaped noise. This suggests Tango's processing techniques are well-suited for maintaining the integrity of the speech signal even in the presence of various noise types. Tango also stands out for reducing the interference of vowels in white noise environments. This is particularly true for open phonemes, which are more vulnerable to interference due to their wider spectral spread. MVDR, on the other hand, outperforms other systems in reducing the interference on vowels in scenarios with speech-shaped noise. FaSnet is overall outperformed by MVDR and Tango, except under speech-shaped noise where it outperforms Tango. Tango's performance appears to depend on the initial amount of interference. The SIR improvement is almost constant for all categories and the output SIR then depends on the phoneme category. MVDR, in contrast, improves the SIR more uniformly, potentially offering a more predictable enhancement outcome, especially when the initial level of noise in vowels are varying.

MVDR also prevails at controlling the presence of artifacts in the estimated speech, which is crucial for the overall perceived quality and intelligibility of the enhanced speech, since a low amount of artifacts implies that the speech signal retains more of its natural characteristics post-estimation. The output SAR obtained with MVDR is once again almost constant regardless of the phoneme category, which could lead to more predictable behavior. The output SAR obtained by Tango depends on the input SIR which makes it potentially less predictable than the MVDR. FaSNest is outperformed by the other systems by a large margin in terms of SAR_out.

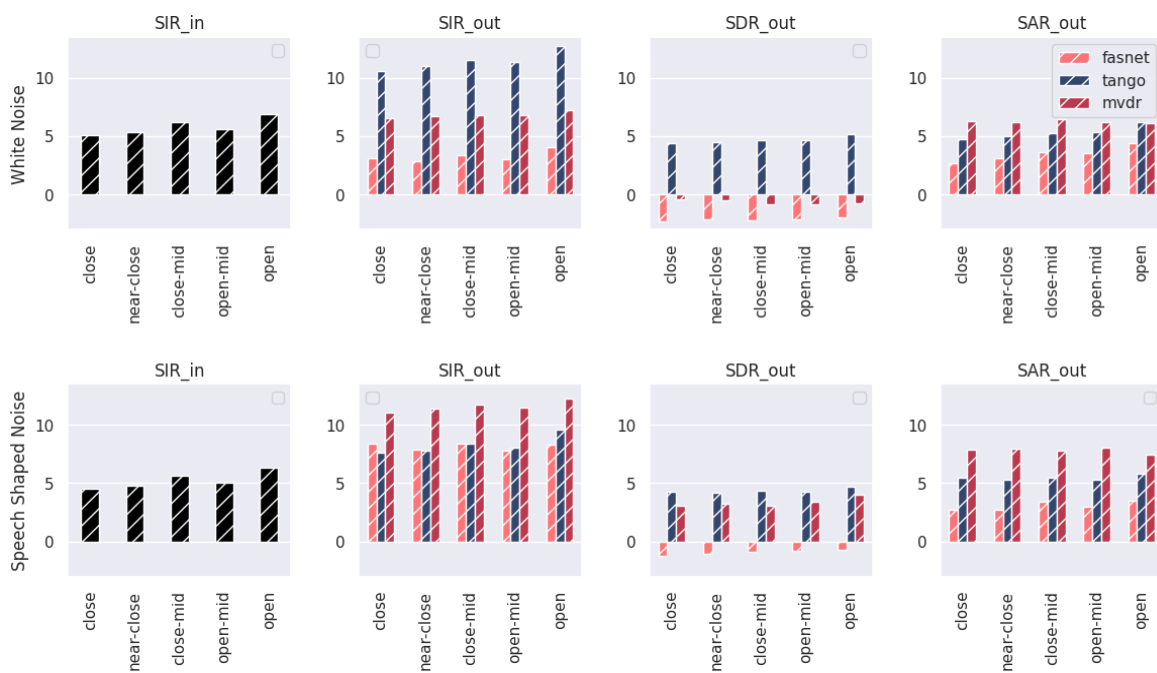

*Figure 17 – Performance of the algorithms on close and open phonemes.*

## Comparison of the algorithms on plosives, fricatives and taps

In this experiment, we compare the performance of the three speech enhancement algorithms across consonant phonemes: plosives, fricatives, taps. As in prior experiments, the analysis is conducted in environments with two types of noise conditions: white noise and speech-shaped noise. The results are presented in Figure 18.

The results reflect the distinct acoustic challenges that plosive, fricative and tap consonants face in the room environment. These consonant types are intrinsically affected by interference due to their articulatory characteristics, with plosives being particularly vulnerable due to the transient nature of sound production, which can be easily masked by environmental noise.

Tango's performance with white noise stands out in its ability to reduce interference, especially in the case of plosives. Despite the challenging initial conditions, Tango manages to enhance plosives' clarity while keeping distortion reasonably low, highlighting its efficacy in dealing with the abrupt and high-intensity nature of plosive sounds. Under speech-shaped noise, Tango does not exhibit any remarkable improvement, yet it maintains its proficiency in interference reduction for plosives. The model's average performance in artifact reduction suggests that while Tango can mitigate some noise elements, there's a trade-off in terms of introducing new artifacts into the signal.

The MVDR exhibits more nuanced results, since we observe a persistence of high distortion levels and low interference improvements when dealing with white noise; however, it consistently reduces interference across all phoneme categories and maintains a high SAR under speech-shaped noise. While Tango performed the best on plosives, the MVDR provides the highest amount of noise reduction on tap phonemes.

Similarly, as on vowels, FasNet fails to reduce interference and introduces a high level of distortion in the presence of white noise. With speech shaped noise, FaSNest exhibits improvement in the SIR. Like for the MVDR, these improvements are larger on tap phonemes than on plosives and fricatives.

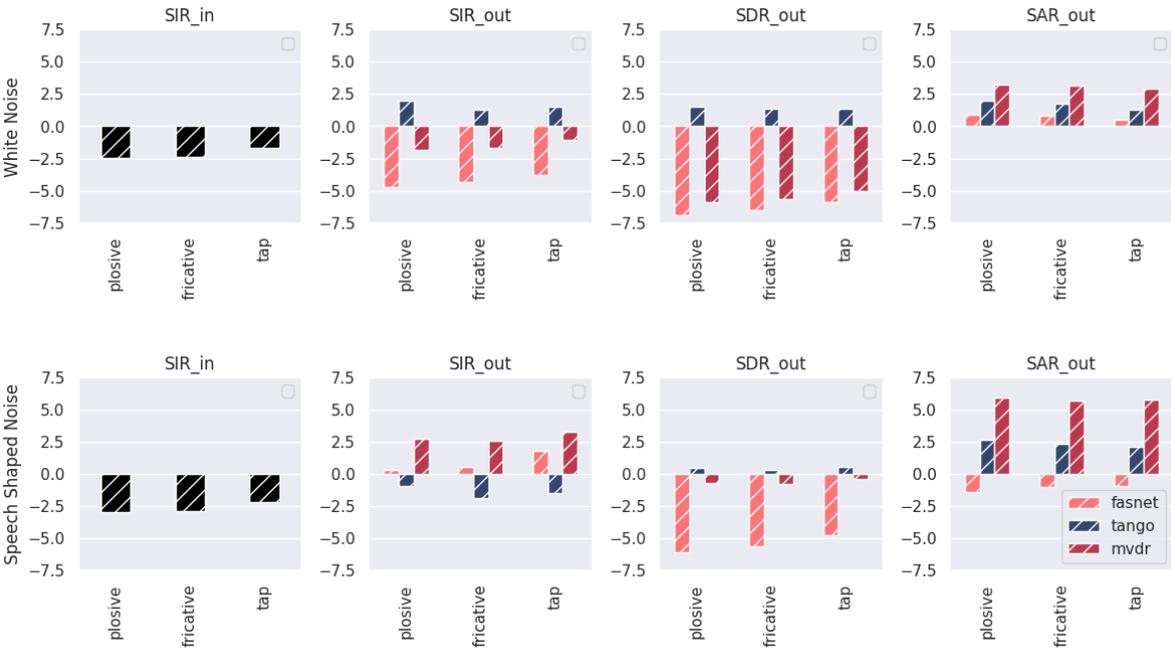

*Figure 18 – Performance of the algorithms on plosives, fricatives and taps.*

## Comparison of the algorithms on approximants and laterals

In this experiment we focus on the performance of the speech enhancement algorithms on approximant and lateral phonemes. The results provide insights into each algorithm's ability to preserve the integrity of these phoneme categories amidst the interference, distortion, and artifacts. The results are presented in Figure 19.

In the presence of white noise, Tango demonstrates proficiency in reducing interference while concurrently keeping distortion levels low, outperforming the other models for both phoneme types. The trend is inverted in the presence of speech-shaped noise where MVDR exhibits the best SIR improvement for both approximants and laterals. Once again, FasNet exhibits a poor performance on white noise and performs on par with Tango in terms of SIR when speech-shaped noise is present. MVDR stands out for its superior control over distortion across both phoneme categories, suggesting its advanced capability to preserve speech quality. Tango and FasNet, while demonstrating lower SARs, successfully maintain a reduced level of artifacts for both types of noise, confirming the algorithms' capabilities in artifact control. It is interesting to note that each model performs consistently across the phoneme class considered on each separate noise condition.

In summary, these experiments show that each algorithm exhibits a different behavior that depends not only on the noise type, but also on the phoneme class.

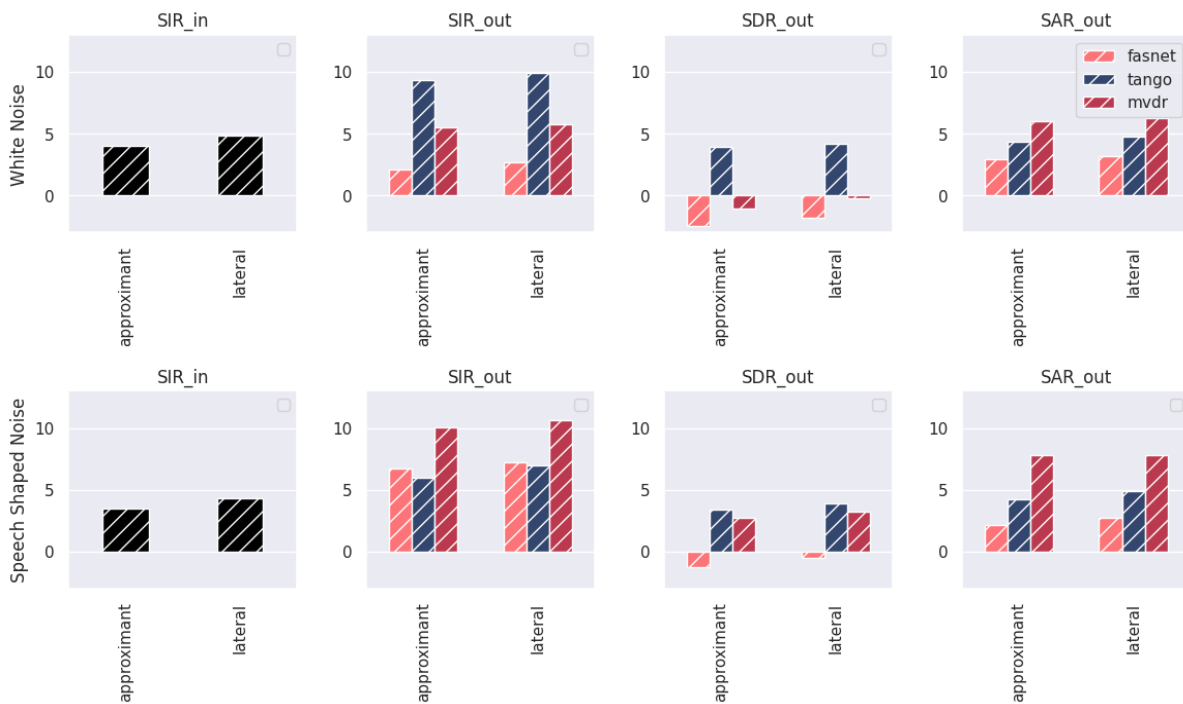

*Figure 19 – Performance of the algorithms on approximant and lateral phonemes.*

## Impact of the level of noise on algorithms' performance

Finally, this experiment aims to understand how each model responds to different levels of background noise across a range of speech sounds, from plosives to open vowels. Figure 20 presents a comparative analysis of the algorithms, evaluating their performance on mixtures with speech-shaped noise.

The plosives, that are characterized by a complete obstruction of the vocal tract, are particularly susceptible to the acoustics interferences, as seen by the consistently lower input SIR compared to the SNR of the mixtures. This suggests that the transient nature of plosives makes them more prone to interference. All models grapple with reducing interference for plosives across scenarios. Even the best performing model (MVDR) only yields a slight SIR improvement. Tango exhibits the worst SIR improvement at –5dB. This could indicate a limit to how much speech enhancement models can counteract the acoustic masking for these rapidly changing phonemes, especially in environments with strong background noise.

In contrast, approximants, closes, and open vowels which involve less abrupt articulatory gestures and more continuous airflow, seem to be easier to enhance even for SNR below 0 dB. This could possibly be due to their more sustained and resonant acoustic signatures that are less easily masked by noise. MVDR yields a noticeable improvement in reducing interference for these phonemes, particularly at 0 dB and 5dB, highlighting its strength in enhancing continuous sounds. Even though this is not as drastic as for the plosive, all the models struggle to improve the SIR when the input SNR is –5 dB. MVDR once again is the approach that performs the best in this case and with speech-shaped noise. Tango and FasNet on the other hand show no improvement for open vowels at –5 dB.

MVDR and Tango maintain a lower level of artifacts in the estimated speech across SNRs from 5 dB to 0 dB for all phonemes. As in previous scenarios, FasNet introduces the largest amount of artifacts regardless of the SNR, as observed in particular for plosives.

This extensive analysis underscores the need for speech enhancement models to be tailored to the acoustic properties of phonemes, considering not only the level of background noise but also the phonetic and articulatory characteristics that define each phoneme's vulnerability to acoustic interference.

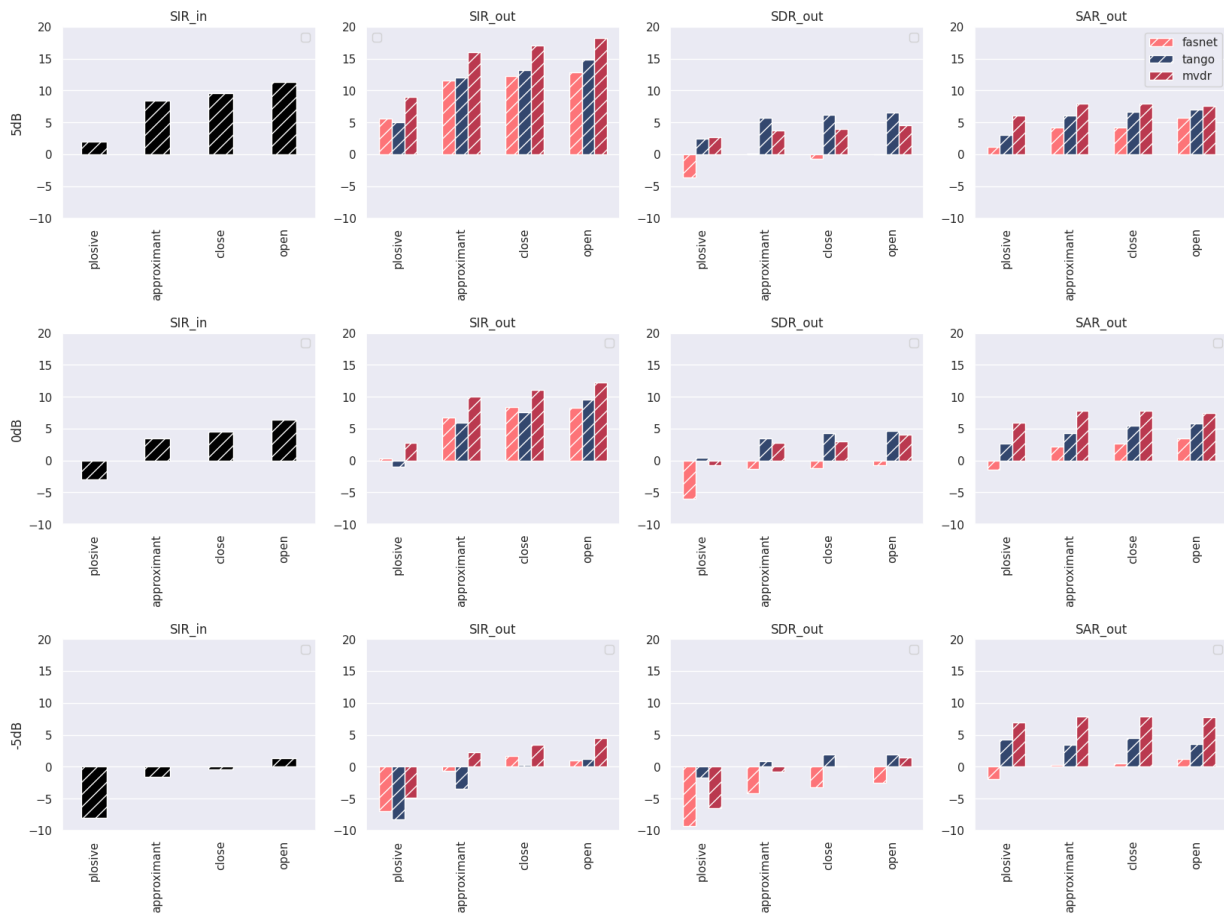

*Figure 20 - Performance of the algorithms on plosives, fricatives, closes and opens with respect to the amount of input noise.*

# Conclusions and Perspectives

In this paper, we conducted a comprehensive evaluation of three state-of-the-art multichannel speech enhancement algorithms (FasNet, MVDR, and Tango), with a particular emphasis on the phoneme-scale analysis. This study revealed that different phonemes are uniquely affected by noise, underlining the limitations of traditional utterance-level evaluations. Specifically, it was found that specific phonemes like plosives are heavily impacted by environmental acoustics, whereas nasals and sibilants show more resistance to noise, especially when it is speech shaped.

Looking forward, this detailed phoneme-scale evaluation framework opens new avenues for tailoring speech enhancement models. It reveals the need for these algorithms to consider the differential impact of noise on various phonemes and adapt accordingly. This research direction can focus on integrating phoneme-specific characteristics into the training of these models, potentially enhancing their effectiveness in real-world noisy environments. Particularly, this could lead to enhance speech intelligibility in real-world scenarios, offering interesting observations for developing more effective, personalized hearing aid technologies.

# Acknowledgments

The authors express gratitude to Chaslav Pavlovic for providing the Portable Hearing Laboratory platform (PHL) used for room acoustic simulations of mixtures.

This work was made with the support of the French National Research Agency, in the framework of the project REFINED "REal-time artificial INtelligence for hEaring aiDs" (ANR-21-CE19-0043). Experiments presented in this paper were partially carried out using the Grid5000 testbed, supported by a scientific interest group hosted by Inria and including CNRS, RENATER and several Universities as well as other organizations (see https://www.grid5000).

## List of Figures

**Figure 20 - Performance of the algorithms on plosives, fricatives, closes and opens with respect to the amount of input noise.**